\documentclass[sigconf]{acmart}
\usepackage{amsmath,amssymb,amsfonts}
\usepackage{algorithmic}
\usepackage{graphicx}
\usepackage{textcomp}
\usepackage{xcolor}
\usepackage{tabularray}

\author{Shahbaz Siddeeq, Zeeshan Rasheed, Malik Abdul Sami, Mahade Hasan, Muhammad Waseem, Jussi Rasku, Mika Saari, Kai-Kristian Kemell, Pekka Abrahamsson}
\author{Henri Terho, Kalle Mäkelä}

\affiliation{%
  \institution{
    \textsuperscript{a}Faculty of Information Technology and Communication Sciences, Tampere University \\
    \textsuperscript{b}Eficode Oy
    }
  \country{Finland}
}
\email{{shahbaz.siddeeq, zeeshan.rasheed, malik.sami, mdmahade.hasan, muhammad.waseem, jussi.rasku, mika.saari, Kai-Kristian.Kemell, pekka.abrahamsson}@tuni.fi}
\email{{henri.terho}@eficode.com, {kalle.makela}@gmail.com}

\begin{document}

\title{Distributed Approach to Haskell Based Applications Refactoring with LLMs Based Multi-Agent Systems \\
% {\footnotesize \textsuperscript{*}Note: Sub-titles are not captured for https://ieeexplore.ieee.org  and
}

% \and
% \and
% \and
% \and
% \and

%\maketitle

\begin{abstract}
We present a large language models (LLMs) based multi-agent system to automate the refactoring of Haskell codebases. The multi-agent system consists of specialized agents performing tasks such as context analysis, refactoring, validation, and testing. Refactoring improvements are using metrics such as cyclomatic complexity, runtime, and memory allocation. Experimental evaluations conducted on Haskell codebases demonstrate improvements in code quality. Cyclomatic complexity was reduced by 13.64\% and 47.06\% in the respective codebases. Memory allocation improved by 4.17\% and 41.73\%, while runtime efficiency increased by up to 50\%. These metrics highlight the system’s ability to optimize Haskell’s functional paradigms while maintaining correctness and scalability. Results show reductions in complexity and performance enhancements across codebases. The integration of LLMs based multi-agent system enables precise task execution and inter-agent collaboration, addressing the challenges of refactoring in functional programming. This approach aims to address the challenges of refactoring functional programming languages through distributed and modular systems.

\end{abstract}

\maketitle

\section{Introduction} \label{sec:intro}
Functional programming languages have long been recognized for their immutability and higher-order functions, which make them suitable for accurate and concurrent processing\cite{hu2015functional}. This foundational strength of functional programming is exemplified in languages like Haskell, which offers a range of features designed for such tasks, including lazy evaluation, type inference, and monadic handling of side effects~\cite{hudak1992gentle, peyton1993imperative}. However, these characteristics also make Haskell based systems challenging to refactor, as modifications must account for dependencies and constructs like monads and type classes \cite{orchard2014embedding, figueroa2021monads}.

Refactoring functional programming languages like Haskell based applications presents a set of challenges~\cite{brown2011expression}. Haskell's functional nature introduces complexity in maintaining code readability, structure, and performance without compromising immutability and higher-level abstractions~\cite{mens2004survey} unlike imperative languages. Traditionally, refactoring efforts in functional languages focus on restructuring code to improve readability and maintainability while ensuring correctness, but the presence of type systems and non-linear evaluation strategies adds difficulty to the process.

 Previous research has explored the use of multi-agent systems in software engineering for tasks such as automated analysis, distributed processing, and iterative improvement~\cite{wooldridge1995intelligent, abdallah2011dynamic, rasheed2024can, rasheed2024autonomous}. These studies have primarily focused on general-purpose programming languages and conventional paradigms. Similarly, the adoption of large language models for code generation, error detection, and basic refactoring has demonstrated potential, but their application to functional programming languages like Haskell remains limited. Existing tools for Haskell refactoring, such as HaRe~\cite{li2005haskell} rely on static or rule-based approaches and do not use the advanced capabilities of LLMs or the collaborative nature of multi-agent system. Multi-agent systems consist of multiple agents that can interact, collaborate, and execute tasks concurrently, which is making them a system for modular and distributed processes \cite{rasheed2024codepori}. In software maintenance, agents have been applied to distributed code analysis and task allocation, with each agent specializing in a function, thereby streamlining processes \cite{rasheed2024large}. Tasks such as code analysis, refactoring, and validation can be distributed among agents to address the complexity in a structured manner by applying multi-agent system principles to Haskell based applications refactoring.

Haskell based application poses challenges while Refactoring due to features such as lazy evaluation, immutability, and type safety. These characteristics demand accuracy in refactoring code while maintaining its functional integrity. Traditional approaches often rely on manual interventions or rule-based systems such as HaRe~\cite{li2005haskell}, which lack the scalability and adaptability required for large and complex codebases. Large Language Models (LLMs) have shown promise in understanding and generating code, yet they struggle with the specific demands of functional programming~\cite{shirafuji2023refactoring, jiang2024survey}. In collaboration of leading Finnish companies working on refactoring of functional programming codebases, this study aims to address these limitations by integrating LLMs based multi-agent system to distribute refactoring tasks.
% and using the strengths of both technologies for efficient and automated Haskell code maintenance. 
The motivation behind this study to fulfills an industry demand for automated refactoring system for large and complex codebases.

\begin{figure}
    \centering
    \includegraphics[width=1\linewidth]{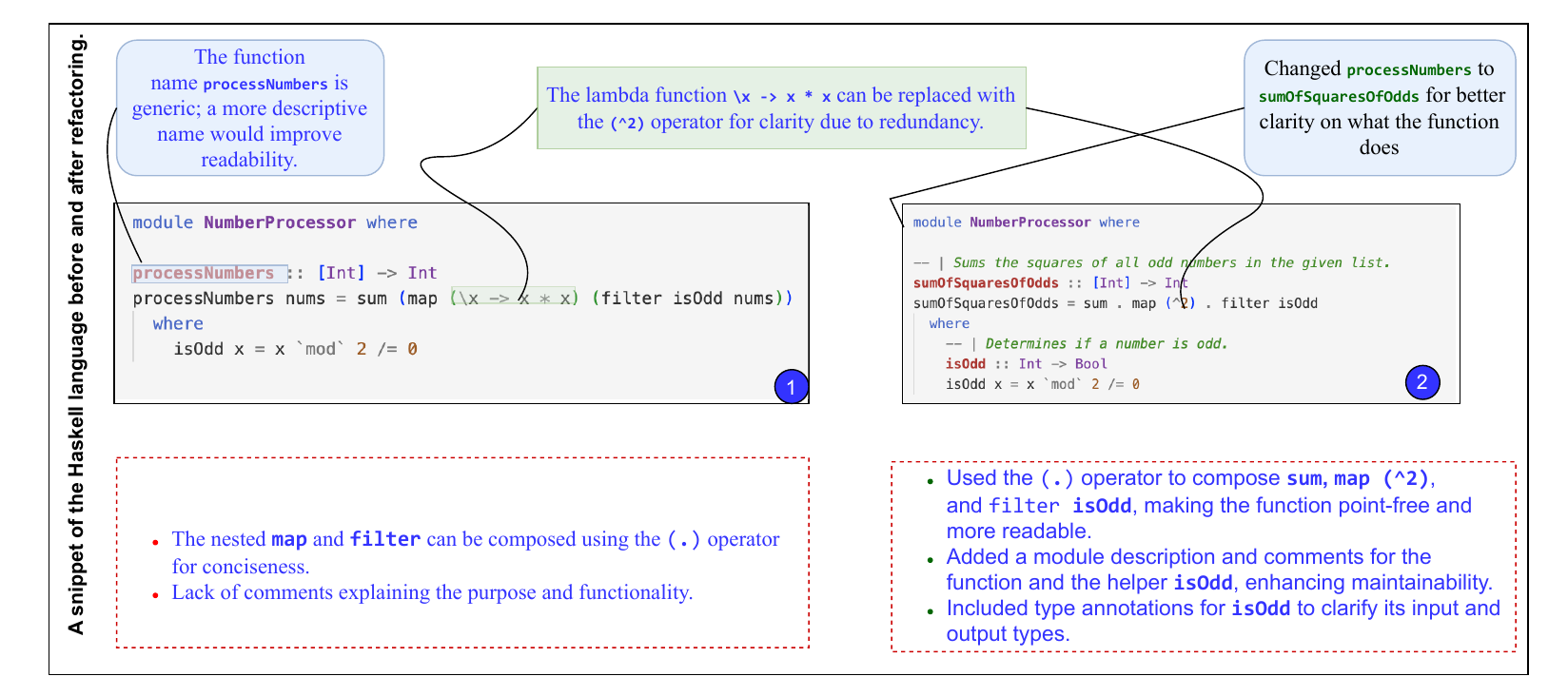}
    \Description{Example: Haskell language before and after refactoring.}
    \caption{
Example: Haskell language before and after refactoring.}
    \label{fig:Example}
\end{figure}

\textbf{Motivating Scenario}:  To contextualize the issues and improvements in Haskell code refactoring, we provide a representative example in Figure \ref{fig:Example}. The example is taken from a hypothetical Haskell project, focusing on a function named \colorbox{lightgray}{\scriptsize{processNumbers}}. This function processes a list of integers, summing the squares of all odd numbers. The figure illustrates the code before and after refactoring, highlighting improvements in readability and conciseness. In the initial version of the code, the function \colorbox{lightgray}{\scriptsize{processNumbers}} lacks descriptive naming and comments, making it difficult to understand its purpose. The lambda function \colorbox{lightgray}{\scriptsize{\textbf{$xx \rightarrow x * x$}}} is redundant and can be simplified. The nested use of map and filter also reduces readability. After refactoring, the function is renamed to \colorbox{lightgray}{\scriptsize{sumOfSquaresOfOdds}} to better reflect its purpose. The lambda function is replaced with the \colorbox{lightgray}{\scriptsize{(*2)}} operator, and the nested operations are composed using the \colorbox{lightgray}{\scriptsize{\textcolor{lightgray}{°}\textbf{.}\textcolor{lightgray}{°}}} operator, making the function point-free and more concise. Comments and type annotations are added to enhance clarity.

This example shows the importance of refactoring in improving code quality. It also highlights common issues in Haskell code, such as unclear function names, redundant expressions, and lack of documentation. The refactored code becomes more accessible to contributors and easier to maintain by addressing these issues.

% To contextualize the refactoring of a Haskell-based system, we provide a representative example in Figure \ref{fig:Example}.

Large Language Models (LLMs) like GPT and Codex have transformed the programming landscape, offering capabilities in code synthesis, error detection, and refactoring. LLMs have shown promise in understanding and generating code structures, facilitating tasks such as automatic code completion and bug detection~\cite{chen2021evaluating, svyatkovskiy2020intellicode} by training on datasets of code. Recently, research has begun to explore the use of LLMs in refactoring tasks~\cite{pomian2024next, dos2015autorefactoring}, where their ability to analyze context and suggest changes has demonstrated effectiveness in code comprehension and enhancement. However, despite these advances, LLMs alone are often insufficient for managing the requirements of functional programming languages, especially in distributed or modular tasks, where multi-agent system approaches may provide support~\cite{guo2024large}.

\textbf{Objective of the study:} The objective of this study is to develop LLMs based multi-agent systems to automate the refactoring of Haskell codebases. The system aims to achieve the following: (i) improve code quality metrics, specifically cyclomatic complexity, runtime efficiency, memory usage, and HLint comparison, and (ii) ensure that the system is adaptable to the characteristics of functional programming paradigms such as lazy evaluation, immutability, and type safety.
% The framework aims to improve cyclomatic complexity, runtime efficiency, and memory usage while ensuring scalability and adaptability to functional programming paradigms.

% In this 
% % work
% \textcolor{blue}{paper}, 
% \textcolor{red}{we propose an approach that integrates large language models based multi-agent systems for refactoring Haskell code. Our system is designed with agents that utilize LLM capabilities for context analysis, refactoring suggestion, code verification, and debugging.}
% \textcolor{blue}{We used an LLM-based multi-agent approach to develop the system for context analysis, refactoring suggestions, code verification, and debugging, providing an efficient and scalable solution for automating the refactoring of Haskell codebases.}

% Each agent is assigned tasks, allowing for modular and efficient handling of refactoring. 
% % For example, the context analysis agent identifies areas of potential improvement, the refactor agent generates code modifications, and the verification agent ensures the refactored code maintains functional integrity. 
% This research aims to introduce an approach for refactoring Haskell codebases
% % ~\cite{ayshwaryalakshmi2013agent}\cite{baumgartner2024ai} 
% by utilizing a multi-agent system framework alongside LLMs \textcolor{blue}{~\cite{ayshwaryalakshmi2013agent}\cite{baumgartner2024ai}}.

\textbf{Contributions}: 
% We introduce a multi-agent approach to automating the refactoring of Haskell codebases by integrating Large Language Models based multi-agent systems. The proposed framework addresses the complexities of functional programming languages, offering improvements in performance, scalability, and adaptability. The system enhances refactoring workflows and sets the groundwork for future advancements in functional programming maintenance by using multi-agent systems for task modularity and LLMs for contextual understanding. 
The primary contributions of this study are:

\begin{itemize}
    \item Development of an LLM-based multi-agent system for refactoring Haskell code.
    \item Evaluation of the developed system against code quality metrics, including cyclomatic complexity, runtime efficiency, memory usage, and comparisons with HLint.
     \item Public release of the developed system, which is made available online~\cite{ShahbazMaSHaskellGitHub2025}, allowing researchers and practitioners to access, replicate, and validate the system. 
     % will add citaion of github link of our code
\end{itemize}

\textbf{Paper organization}: Section \ref{sec:rw} describes the related work; Section \ref{sec:rm} presents the research method; Section \ref{sec:results} describes the results and evaluation; Section \ref{sec:discussion} discusses this work with limitations and future research directions; Section \ref{sec:threat} describes threats to validity; Section \ref{sec:conclusions} concludes this work.

\section{Related Work} \label{sec:rw}
\subsection{Functional Programming Languages and Haskell Refactoring}

Functional programming languages like Haskell present challenges due to their emphasis on immutability, lazy evaluation, and higher-order functions~\cite{bragilevsky2021haskell}. These
challenges complicate the refactoring process, especially when changes need to preserve program semantics~\cite{thompson2013refactoring}. Most research on refactoring functional programming based applications primarily focuses on traditional methods. Mens and Tourwé~\cite{mens2004survey} conducted a comprehensive survey highlighting refactoring patterns applicable across programming paradigms, including managing side effects and optimizing type safety. Further explored the challenges of refactoring functional programs~\cite{abdallah2011dynamic}, proposing techniques tailored for functional codebases to maintain semantic integrity while enhancing readability and performance.

Despite efforts, there is still limited research on automated refactoring in Haskell and most methods depending on rule-based systems or requiring manual intervention. These limitations highlight the need for automated systems that can understand dependencies within Haskell code and suggest effective refactoring solutions. Our research seeks to bridge this gap by introducing a multi-agent system capable of performing distributed and automated refactoring tasks within a Haskell codebase.

\subsection{Multi-Agent Systems in Software Refactoring}

Multi-Agent Systems have evolved since their inception in the mid-1990s, when foundational ideas emphasized distributed collaboration and task specialization using rule-based approaches~\cite{wadler1992essence}. Between 2010-2020, the multi-agent system allows for multiple agents to specialize in different tasks and collaborate to achieve a common goal, which makes it useful for complex codebases requiring iterative improvement~\cite{abdallah2011dynamic}. AyshwaryaLakshmi et al.~\cite{ayshwaryalakshmi2013agent} explored the use of multi-agent systems in distributed code refactoring, demonstrating how autonomous agents can collaboratively handle large-scale refactoring tasks with improved efficiency and accuracy. After 2021, unlike earlier systems, LLMs empower agents to interpret complex domains, such as Haskell's functional programming features, and automate tasks like refactoring and debugging with unprecedented scalability~\cite{chen2021evaluating}. This combination bridges theoretical concepts with practical marking a leap in multi-agent system capabilities and redefining their role in modern software engineering.

% Multi-agent systems have shown promise in distributed and autonomous decision-making environments. In software engineering, multi-agent system has been applied to tasks requiring parallelized efforts, such as code analysis, refactoring, and maintenance~\cite{wooldridge1995intelligent}. The multi-agent system allows for multiple agents to specialize in different tasks and collaborate to achieve a common goal, which makes it useful for complex codebases requiring iterative improvement~\cite{abdallah2011dynamic}. AyshwaryaLakshmi et al.~\cite{ayshwaryalakshmi2013agent} explored the use of multi-agent systems in distributed code refactoring, demonstrating how autonomous agents can collaboratively handle large-scale refactoring tasks with improved efficiency and accuracy. By distributing refactoring responsibilities, multi-agent system enables greater flexibility and modularity in tackling complex code structures.

For Haskell refactoring, multi-agent system can be advantageous, as agents can be assigned tasks like context analysis, refactoring suggestion, and validation \cite{dos2015autorefactoring}. Each agent’s specialization allows for more effective handling of Haskell’s unique functional constructs. Our research uses multi-agent system principles to distribute Haskell refactoring tasks across multiple agents~\cite{guo2024large}, each enhanced by LLM capabilities, which collectively provide a scalable approach for functional programming language maintenance.

\subsection{LLMs for Code Generation and Refactoring}

The rise of large language models (LLMs) has revolutionized automated code generation and refactoring such as OpenAI’s GPT series and Codex. These models, trained on vast datasets of code, have demonstrated proficiency in tasks ranging from code completion to error detection and refactoring suggestions~\cite{chen2021evaluating}. Svyatkovskiy et al.~\cite{svyatkovskiy2020intellicode} introduced IntelliCode Compose reducing the time developers spend on routine coding tasks which is a transformer-based system that uses LLMs to assist in code generation. Recent work~\cite{white2024chatgpt}\cite{hou2024large} has shown that LLMs can not only generate syntactically correct code but also adapt to specific programming paradigms, making them valuable assets for software refactoring.

However, while LLMs are effective at generating and analyzing code, they often lack the domain-specific insights required for functional programming languages. Additionally, in distributed refactoring tasks, LLMs alone may be insufficient due to limitations in handling complex, modular tasks autonomously. Our work addresses these limitations by integrating LLMs into a multi-agent system, where the LLMs augment the capabilities of each agent in tasks like code analysis, refactoring suggestion, and debugging. This integration allows for a more contextually aware, automated refactoring process tailored to the complexities of Haskell.

\subsection{Combining Multi-Agent Systems and LLMs for Haskell Refactoring}

The LLMs based multi-agent systems for code refactoring is an approach that has not been extensively explored in the literature. Recently proposed~\cite{baumgartner2024ai}\cite{xi2023rise} a multi-agent learning system for automatic code refactoring, which demonstrated improvements in handling distributed code modifications. However, their approach was primarily focused on imperative programming languages and did not explore applications in functional programming. Similarly, discussed~\cite{ayshwaryalakshmi2013agent}\cite{huang2024levels} a multi-agent approach to software maintenance but their work did not consider the challenges associated with functional languages like Haskell.

Our research expands on these prior works by introducing a multi-agent system~\cite{hua2023war} specifically designed for functional programming refactoring~\cite{dos2015autorefactoring}, enhanced by the contextual analysis and code generation capabilities of LLMs. By assigning each agent a specialized task and utilizing LLMs for deeper language comprehension, our approach aims to streamline Haskell refactoring ~\cite{thompson2013refactoring} in a scalable, autonomous manner. This combination of multi-agent systems and LLMs provides a solution to the challenges of functional programming language maintenance and extends the capabilities of both multi-agent systems and LLMs in software engineering~\cite{cheng2024exploring}.

\section{Research Method} \label{sec:rm}
\begin{figure*}[h]
    \centering
    
    \includegraphics[width=\textwidth, clip, trim=0 100 0 0]{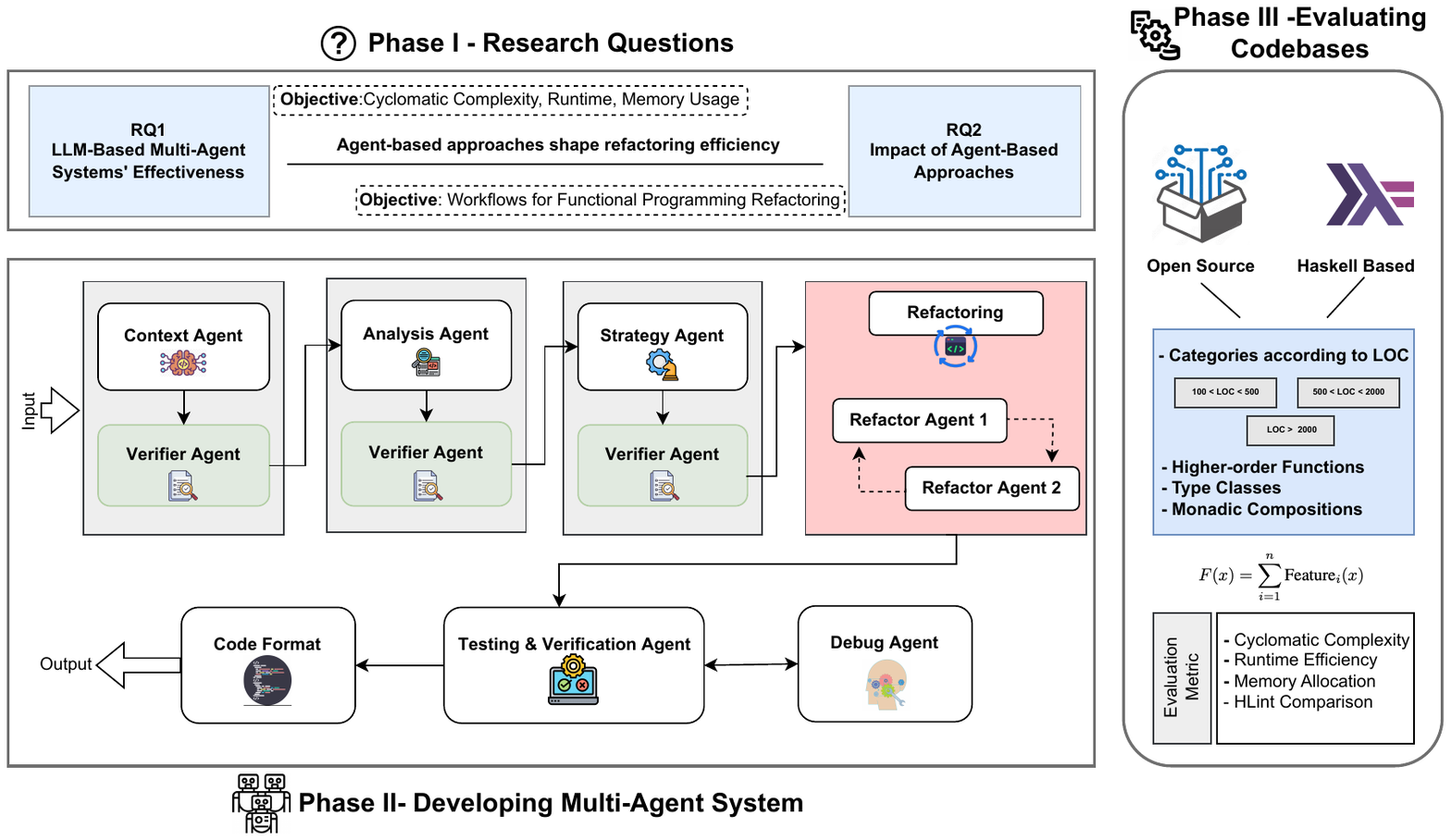}
    \Description{Research Methodology}
    \caption{Research Methodology}
    \label{fig:RM}
\end{figure*}

The methodology employed for this study is divided into three
phases, elaborated below and illustrated in Figure \ref{fig:RM}

\subsection{Phase I - Research Questions}
Considering our research objective, we formulated the following two RQs.
  \begin{itemize}
      \item \textbf{RQ1}: How effectively can Large Language Models (LLMs) based multi-agent systems improve Haskell codebases in terms of cyclomatic complexity, runtime, and memory usage? The \textbf{objective} of this RQ is to evaluate the effectiveness of LLM-based multi-agent systems in automating Haskell code refactoring, specifically focusing on reducing cyclomatic complexity, improving runtime performance, and optimizing memory usage
      \item \textbf{RQ2}: What is the impact of using agent-based approaches on refactoring workflows for functional programming languages?The \textbf{objective} of this RQ is to assess the flexibility of a multi-agent system in addressing the challenges of refactoring Haskell code, particularly focusing on aspects such as immutability, lazy evaluation, and type safety.
 \end{itemize}

% \begin{figure*}[h]
%     \centering
%     \includegraphics[width=\textwidth]{Multi_Agent_System.png}
%     \caption{Multi-Agent System}
%     \label{fig:MAS}
% \end{figure*}
\subsection{Phase II- Developing Multi-Agent System}
% Refactoring legacy codebases is a challenging task in software engineering, particularly for functional programming languages like Haskell. Addressing inefficiencies, improving modularity, and ensuring the follow-up of best practices requires a systematic approach. 
%The proposed multi-agent system where agents perform specific roles, ensuring an effective refactoring workflow. Figure~\ref{fig:MAS} is showing the system incorporates a sequence of specialized agents, each contributing to an automated refactoring pipeline. The design ensures scalability and efficiency. The agents operate in a feedback loop, where outputs from one agent are validated or extended by the next. The process is based on functional programming principles, particularly for Haskell codebases.

We developed an LLM-based multi-agent system to automate the refactoring process for Haskell codebases. The system consists of specialized LLM-based agents, each performing a specific role to ensure an efficient and organized workflow. The agents operate sequentially, with the output of one agent being validated or refined by the next. Figure~\ref{fig:RM} illustrates the key components of the system and their interactions. Below, we provide a brief overview of each agent along with there responsibilities.
\begin{itemize}
    \item \textbf{Code Context Agent:} The Code Context Agent is responsible for analyzing the codebase and providing a clear structural overview. It begins by parsing the code to identify key components, such as modules, functions, and their interdependencies. Next, it generates visual and logical representations, such as control flow diagrams, to illustrate the architecture of the codebase. These representations provide essential context for subsequent agents, ensuring that their analysis and refactoring tasks are precise and effective. 
    
    %The Code Context Agent is responsible for parsing and structuring the codebase. It identifies components, such as modules, functions, and their interdependencies, and generates representations like control flow diagrams. These outputs enable subsequent agents to operate with a clear understanding of the codebase's architecture. By providing this context, the Code Context Agent ensures that analysis and refactoring tasks are well-defined and effective.
    
    %In software engineering, creating structural representations is essential for tasks like modularization and dependency management. Research on software visualization and reflexion models, such as by Murphy et al.~\cite{murphy1995software}, highlights the need to bridge designs with implementation for effective software maintenance and refactoring~\cite{murphy1995software}\cite{spinellis2006code}.

    \item \textbf{Context Verifier Agent}: The Verifier Agent ensures the accuracy and reliability of the structural data provided by the Code Context Agent. It begins by cross-validating the identified dependencies and modular structures to confirm that they accurately reflect the codebase's implementation. By resolving ambiguities and inconsistencies, the Verifier Agent eliminates potential issues that could lead to inefficiencies or errors in later analysis and refactoring stages.
    %The Code Context Verifier Agent verifies the structural data provided by the Code Context Agent. It cross-validates dependencies and modular structures to confirm that the codebase's context matches the implementation. This agent removes ambiguities that could lead to inefficiencies or errors in subsequent analysis and refactoring phases.
    
    %The role of verification in structural analysis is supported by static analysis research. Works like Clarke et al.’s Model Checking highlight the need for validation to ensure correctness in software representation~\cite{clarke1999model}. This step is essential for maintaining automated workflows.
        
   % \item \textbf{Refactoring Analysis Agent:} The Code Analysis Agent performs detailed analysis of the codebase to identify inefficiencies and bottlenecks. It calculates key metrics, such as Cyclomatic Complexity (CC), using McCabe’s formula: 
    %\[CC = E - N + 2P\] 
    %Where~\textbf{\textit{E}} represents the number of edges in the control flow graph, \textbf{\textit{N}} is the number of nodes, and \textbf{\textit{P}} denotes the number of connected components or entry points. This analysis highlights areas of high complexity or poor modularity, forming the basis for refactoring strategies.
    \item \textbf{Refactoring Analysis Agent:} The Refactoring Analysis Agent conducts a thorough evaluation of the codebase to identify inefficiencies and areas requiring improvement. It calculates key metrics, such as Cyclomatic Complexity (CC), using McCabe’s formula:
\[
CC = E - N + 2P
\]
where \textbf{\textit{E}} represents the number of edges in the control flow graph, \textbf{\textit{N}} is the number of nodes, and \textbf{\textit{P}} denotes the number of connected components or entry points. By analyzing these metrics, the agent identifies regions of high complexity and poor modularity, which are prioritized for refactoring. This detailed analysis provides the foundation for developing effective refactoring strategies, ensuring improved code structure and maintainability.

   % Mens and Tourwé’s work on software refactoring demonstrates how metrics like Cyclomatic Complexity support maintaining and improving codebases~\cite{mens2004survey}\cite{mccabe1976complexity}. This agent applies these insights to identify areas for improvement.

    \item \textbf{Analysis verifier Agent:} 
    The Analysis Verifier Agent validates the findings of the Refactoring Analysis Agent to ensure accuracy. It verifies metrics like Cyclomatic Complexity and confirms that areas marked for improvement correspond to actual inefficiencies. This agent eliminates false positives in the analysis and ensures the reliability of the data that guide subsequent refactoring efforts.
    % The Analysis Verifier Agent validates the findings of the Code Analysis Agent to ensure accuracy. It cross-checks Cyclomatic Complexity calculations and confirms that inefficiencies reflect performance issues. This validation step strengthens the analysis, providing a solid foundation for refactoring.

    % Error detection and validation systems for software metrics play a critical role in maintaining software reliability as highlighted by Spinellis~\cite{spinellis2006code} in Code Quality. The Analysis Verifier Agent ensures that refactoring strategies are based on precise and verified data.
    
    \item \textbf{Refactoring Strategy Agent:} 
    The Refactoring Strategy Agent formulates refactoring strategies based on the analysis provided by the Refactoring Analysis Agent. It recommends modular decompositions, function simplifications, and optimizations of data structure to improve performance. This agent ensures that all strategies are customized to the characteristics of the codebase.
    
    % The Refactoring Strategy Agent formulates refactoring strategies based on analysis. It suggests decomposition, function optimizations, and data structure refinements to improve code maintainability and performance.  The Refactoring Strategy Agent ensures these strategies follow practices relevant to the codebase.

    % Gamma et al.'s~\cite{gamma1995design} Design Patterns highlights principles of modularization and software maintainability that inform the Refactoring Strategy Agent approach. This agent converts analytical insights into improvement plans.

    \item \textbf{Strategy Verifier Agent:} 
    The Strategy Verifier Agent evaluates the feasibility of the strategies proposed by the Refactoring Strategy Agent. It ensures that each strategy aligns with findings and and refactoring goals. The agent validates that proposed changes, such as breaking down functions or reorganizing modules, enhance code quality without introducing new issues.
    
    % The Refactoring Strategy Verifier Agent confirms that the strategies from the Refactoring Strategy Agent are feasible and aligned with the findings and refactoring goals. It reviews the proposed decompositions, function restructurings, and data structure optimizations to ensure they reduce complexity, enhance performance, and improve maintainability. For example, the Refactoring Strategy Verifier Agent checks that breaking down a function into smaller units reduces interdependencies without introducing fragmentation.
    
    \item \textbf{Refactor Code Agents:} Refactor Agents are responsible for implementing refactoring strategies while preserving the correctness and functionality of the codebase.

    \begin{itemize}
        \item \textbf{Refactor Agent 1:} Initiates the refactoring process. It focuses on simplifying complex logic, removing redundancies, and improving code readability. This agent applies straightforward refactoring rules, such as consolidating duplicate code, breaking down large functions, and renaming variables for clarity. The Refactor 1 Agent lays the groundwork for subsequent phases by establishing a cleaner code structure.
        \item \textbf{Refactor Agent 2:} Building on the groundwork set by the Refactor Agent 1, Refactor Agent 2 conducts advanced refactoring tasks. It focuses on optimizing performance, and enforcing coding standards. This agent refines the code further by addressing inefficiencies, reorganizing modules. The Refactor Agent 2 restructures the code into a strong form.
    \end{itemize}

    % The Refactor Code Agent applies the refactoring strategies approved by the Strategy Verifier Agent. It executes changes to the codebase, including restructuring functions, optimizing loops, and refining data structures. This agent ensures that the code's behavior remains same while implementing improvements to its design and performance.
    
    % The Refactor Code Agent implements the refactoring strategies devised by the Refactoring Strategy Agent. It restructures functions, improves recursive calls, and updates data structures while preserving the code’s correctness. The Refactor Code Agent converts improvements into practical changes.
    
    % Refactoring tools like Eclipse’s refactoring engine show how implementation affects software quality~\cite{pomian2024next}\cite{fowler2018refactoring}. The Refactor Code Agent automates these tasks to improve outcomes.

    % \item \textbf{Refactor Code Verifier Agent:} 
    
    % The Refactored Code Verifier confirms that the refactored code follows the proposed strategy and maintains correctness. It validates outputs, checks adherence to modularization principles, and benchmarks performance against metrics to verify improvements.
    
    % Research on static analysis and testing tools, such as QuickCheck for Haskell, highlights the need for validation in refactoring workflows~\cite{claessen2011quickcheck}\cite{beck2022test}. The Refactored Code Verifier ensures the refactored code works as intended.

    \item \textbf{Testing and Validation Agent:} The Testing and Validation Agent tests the refactored code to ensure that it meets all functional and performance requirements. This agent ensures that the code is stable and ready for deployment by running test suites, including unit, integration, and system tests. Validation reports summarize the test results and highlight any issues.
    
    % The Testing and Validation Agent tests the refactored code to ensure it meets functionality, performance, and scalability requirements. By running unit tests, integration tests, and performance benchmarks, the Testing and Validation Agent verifies the refactored code is ready for deployment. Validation reports provide transparency and insights.
    
    % Frameworks like QuickCheck and research on test-driven development by Beck inform the Testing and Validation Agent’s methodology~\cite{claessen2011quickcheck}\cite{beck2022test}. This agent ensures the final product meets quality standards.

    \item \textbf{Debug Agent:} The Debug Agent identifies and resolves issues introduced during the refactoring process. It ensures the code operates as intended by analyzing runtime errors, and inspecting control flows. This agent is essential to maintain the reliability of the refactored codebase.
    
    % The Debug Agent identifies and fixes issues introduced during refactoring, such as regressions or inefficiencies. It uses debugging techniques like control flow analysis and runtime monitoring to locate and resolve anomalies. This agent prevents the refactoring process from compromising the codebase.

    % Systematic debugging methodologies, such as those discussed by Zeller in Why Programs Fail, highlight the importance of integrating automated debugging into refactoring pipelines~\cite{zeller2009programs}\cite{jones2002visualization}. The DA ensures the system’s stability throughout the workflow.
    
\end{itemize}

\subsection{Phase III -Evaluating Codebases}
The codebases evaluation process involves three critical steps: deciding the evaluation criteria, selecting representative codebases, and executing the refactoring process while measuring performance.
% using both qualitative and quantitative metrics.
\begin{itemize}
    \item \textbf{Deciding Codebases:} The selection of Haskell codebases for evaluation is guided by a combination of quantitative and qualitative criteria. Quantitatively, the size of the codebase, measured in lines of code (LOC), is a key parameter. Codebases are grouped into three categories: small ($100 \leq \text{LOC} < 500$), medium ($500 \leq \text{LOC} < 2000$), and large ($\text{LOC} \geq 2000$). This classification ensures that the system's scalability and performance are tested across various complexities. Additionally, the functional programming constructs present in the codebase, such as higher-order functions, type classes, and monadic compositions, are evaluated using a feature count metric:
    
    \begin{equation}
    F(x) = \sum_{i=1}^{n} \text{Feature}_i(x)
    \end{equation}

    where $\text{Feature}_i$ represents individual functional constructs. The inclusion of application domains, such as web development, data processing, and compilers, adds strength to the evaluation. For example, the Yesod web framework and the Pandoc text-processing library are exemplary Haskell projects that meet these criteria.

    \item \textbf{Selecting Codebases:} To identify suitable codebases, open-source repositories were sourced from platforms such as GitHub and Hackage. Projects were filtered using advanced search queries, such as \colorbox{lightgray}{\scriptsize{language:Haskell}} \colorbox{lightgray}{\scriptsize{$stars:>50$}} \colorbox{lightgray}{\scriptsize{$size:<2000$}}, to ensure active maintenance and community relevance. 
    % Selected projects included the Lens library, which demonstrates sophisticated type-class usage, and the Stack tool, a build system emphasizing modularity. 
    Each project was manually inspected to ensure the presence of key functional programming constructs and the compliance with coding standards. 
    % For instance, the Lens library was chosen for its extensive use of advanced type features like generalized algebraic data types (GADTs) and lenses. 
    To further validate the selection, tools like HLint were applied to assess pre-refactoring code quality by quantifying code smells, and the results informed the choice of codebases.

    \item \textbf{Refactoring Codebases:} The selected codebases were processed through the multi-agent system in a controlled environment. Initially, baseline metrics were recorded including cyclomatic complexity ($C_{\text{pre}}$), Runtime Efficiency ($T_{\text{pre}}$), and Memory Allocation ($M_{\text{pre}}$).
    %, code smells ($Q_{\text{pre}}$), and LOC ($L_{\text{pre}}$). 
    The multi-agent workflow began with the Context Agent, which  analyzed the codebase. Next, the Refactoring Agent applied transformations to reduce $C_{\text{pre}}$, and memory allocation, focusing on simplifying recursive functions and improving code readability. The Validation Agent ensured that refactored code maintained functional equivalence. For example, the refactoring of the Pandoc library's monadic parsing functions led to a 15\% reduction in cyclomatic complexity, as measured by post-refactoring metrics ($C_{\text{post}}$). 
    % Execution time ($T_{\text{refactor}}$) and memory usage ($M_{\text{refactor}}$) during the refactoring process were also recorded, demonstrating a negligible overhead, which highlights the system's efficiency. 
    These results were benchmarked against manual refactoring processes and existing tools like HLint to validate the improvements.
\end{itemize}

% \subsection{Workflow and Collaboration}

\begin{table*}[ht!]
\centering
\caption{Pre-Refactor and Post-Refactor Results for Codebase A and B}
\label{tab:summary1}
\def\arraystretch{2}%
\begin{tabular}{l|ll|ll}
\textbf{}                           & \multicolumn{2}{c|}{\textbf{Pre-Refactor}}                                                                                                          & \multicolumn{2}{c}{\textbf{Post-Refactor}}                                                                                                           \\ \hline
\textbf{Codebases}                  & Codebase A                                                              & Codebase B                                                                & Codebase A                                                               & Codebase B                                                                \\ \hline
\textbf{Cyclomatic Complexity (CC)} & 22                                                                      & 17                                                                        & 19                                                                       & 09                                                                        \\ \hline
\textbf{Runtime Efficiency}                    & \begin{tabular}[c]{@{}l@{}}0.01 secs \\ (4 ticks @ 1000 us)\end{tabular}     & \begin{tabular}[c]{@{}l@{}}0.01 secs \\ (13 ticks @ 1000 us)\end{tabular} & \begin{tabular}[c]{@{}l@{}}0.01 secs \\ (2 ticks @ 1000 us)\end{tabular}      & \begin{tabular}[c]{@{}l@{}}0.01 secs \\ (12 ticks @ 1000 us)\end{tabular} \\ \hline
\textbf{Memory Allocation}          & \begin{tabular}[c]{@{}l@{}}300,496 bytes \\ ($\sim$0.3 MB)\end{tabular} & \begin{tabular}[c]{@{}l@{}}2,059,288 bytes \\ (2 MB)\end{tabular}         & \begin{tabular}[c]{@{}l@{}}287,952 bytes \\ ($\sim$0.28 MB)\end{tabular} & \begin{tabular}[c]{@{}l@{}}1,200,040 bytes \\ (1.2 MB)\end{tabular}       \\ \hline
\textbf{HLint Comparison}           & 2 hints                                                                 & 2 hints                                                                   & 3 hints                                                                  & 1 hints                                                                  
\end{tabular}
\end{table*}

\subsection{Evaluation}
The effectiveness of the multi-agent system is evaluated using state-of-the-art benchmarks and metrics specific to functional programming and software engineering. This ensures that the refactoring outputs meet established standards for quality, performance, and maintainability.
\begin{itemize}
  \item \textbf{Code Complexity Reduction:} Cyclomatic complexity and structural dependencies are measured pre-refactoring and post-refactoring using established software metrics~\cite{mccabe1976complexity}. This provides a quantitative measure of simplification.

    To measure the complexity of the codebases, we utilized McCabe's Cyclomatic Complexity metric, which quantifies the number of linearly independent paths through a program. Cyclomatic Complexity values were calculated using standard static analysis tools before and after the refactoring process. Higher values indicate increased complexity, while reductions suggest improved maintainability and readability of the code. The effectiveness of complexity reduction was benchmarked against prior studies.

    Cyclomatic Complexity is a software metric that measures the number of linearly independent paths in a program. It indicates the program's complexity and maintainability. The formula for calculating CC is:
    \begin{equation}
        CC = E - N + 2P
    \end{equation}
    % \[CC = E - N + 2P\]
    where:
    \begin{itemize}
        \item \(E\): Number of edges in the control flow graph,
        \item \(N\): Number of nodes in the control flow graph,
        \item \(P\): Number of connected components (typically 1 for a single program).
    \end{itemize}
    
    For multiple functions, the total complexity is:
    \begin{equation}
         CC_{\text{Total}} = \sum_{i=1}^{n} CC_i
    \end{equation}
    % \[CC_{\text{Total}} = \sum_{i=1}^{n} CC_i\]

    Cyclomatic Complexity is used to evaluate the maintainability of code. High CC values indicate code that is harder to understand, test, and maintain. By reducing CC through refactoring, code clarity and efficiency are improved. 
    
    \item \textbf{Performance Benchmarks:} Runtime and memory usage are evaluated using GHC profiling tools. This aligns with prior studies on performance optimization in Haskell~\cite{gill2009worker}.
    
    Runtime refers to the execution time of a program, often measured in ticks or seconds. Memory usage refers to the total bytes allocated during execution. These metrics are critical for assessing the efficiency of a program.

    Using the Glasgow Haskell Compiler (GHC) profiling tools (+RTS options), the runtime and memory performance of the codebases were evaluated.

    \item \textbf{Comparison with HLint:} HLint is a static code analysis tool for Haskell that suggests stylistic improvements to enhance code readability and maintainability.

    HLint was used to evaluate the stylistic improvements in the refactored codebases.
        
\end{itemize}

Multi-agent system is benchmarked against traditional refactoring tools for Haskell, such as HLint and GHC plugins. Performance gains, complexity reduction, and maintainability improvements are quantitatively and qualitatively analyzed to demonstrate the advantages of the proposed system.

This evaluation integrates automated testing frameworks~\cite{prechelt2000empirical} and uses previous work on program synthesis and optimization in functional programming~\cite{chen2021evaluating}. These benchmarks provide validation for the multi-agent system's outputs.

\section{Results} \label{sec:results}

This section presents the evaluation results obtained from analyzing and refactoring Codebase A and Codebase B. Metrics such as cyclomatic complexity, runtime, memory usage, and HLint suggestions were used to quantify the improvements. The results are illustrated using figures and summarized in tables to provide a view of the refactoring process's impact.

% \subsection{Cyclomatic Complexity Reduction}

\subsection{LLM-Based Refactoring Efficiency (RQ1)}

\textbf{Cyclomatic Complexity Reduction:} Cyclomatic Complexity (CC) measures the complexity of a codebase by evaluating the number of linearly independent paths through the program. Higher CC values often indicate a more complex and harder-to-maintain codebase. We observed reductions in CC for both codebases by applying the refactoring techniques discussed in the methodology.

\begin{figure}[h]
\centering
\includegraphics[scale=0.27]{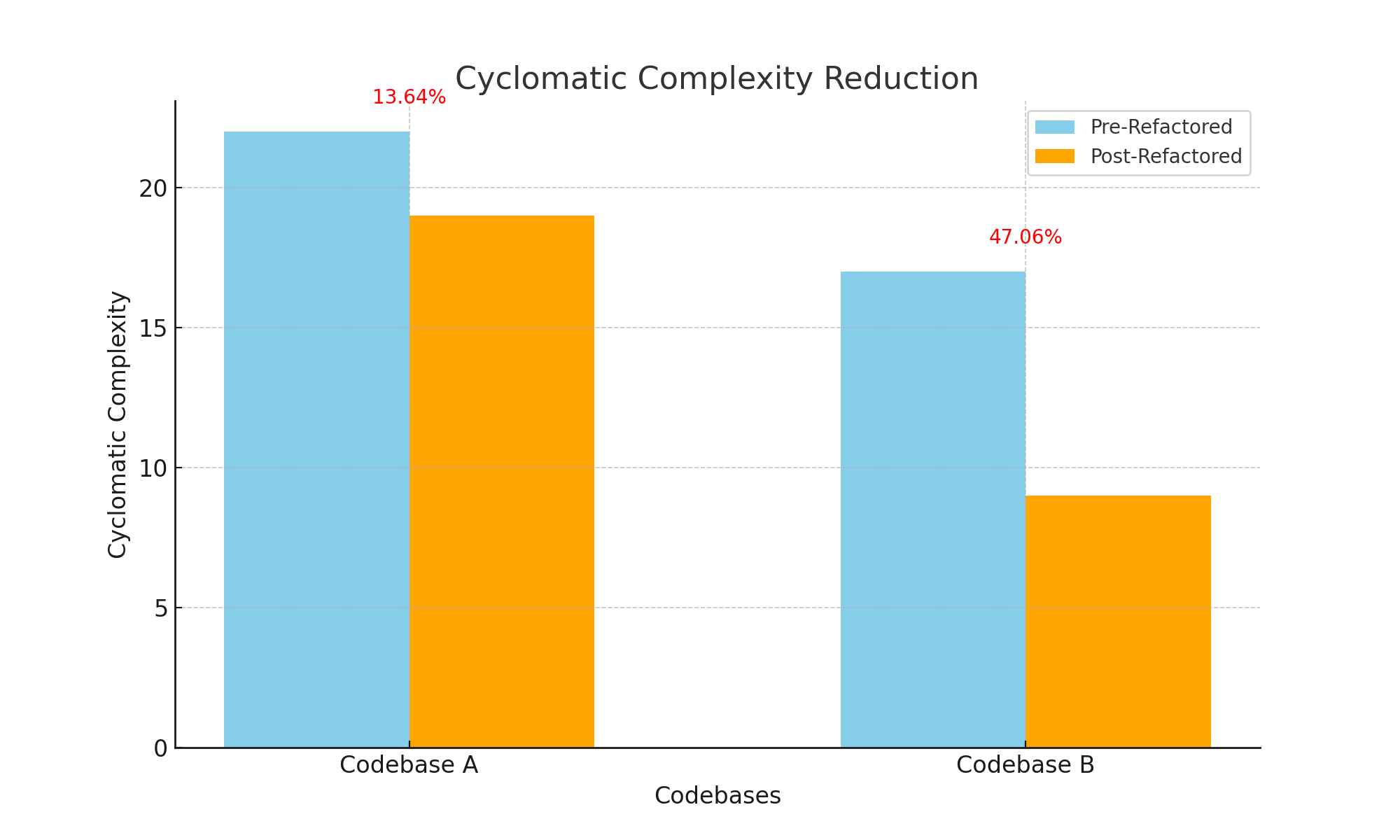}
\Description{Cyclomatic Complexity Reduction for Codebase A and B}
\caption{Cyclomatic Complexity Reduction for Codebase A and B}
\label{fig:cc_reduction}
\end{figure}

\begin{itemize}
    \item \textbf{Codebase A:} The CC decreased by 13.64\%, from 22 to 19. This reflects an improvement achieved by reducing the number of conditional branches and simplifying the control flow.
    \item \textbf{Codebase B:} The CC decreased by 47.06\%, from 17 to 9, indicating simplification of the code structure, which improved readability.
\end{itemize}

Figure~\ref{fig:cc_reduction} illustrates the CC reductions, and the detailed values are summarized in Table~\ref{tab:summary1}.

% \begin{figure}
%     \centering
%     \includegraphics[width=1.0]{bar_graph_cc_reduction.png}
%     \caption{Cyclomatic Complexity Reduction for Codebase A and B}
%     \label{fig:cc_reduction}
% \end{figure}

% \subsection{Runtime and Memory Usage Optimization}

\textbf{Runtime and Memory Usage Optimization:} Optimized runtime and memory usage are critical for improving software performance. Using the GHC profiling tools, we measured runtime (ticks) and memory allocation (bytes) before and after refactoring. The results are as follows:

\begin{itemize}
    \item \textbf{Codebase A:}
    \begin{itemize}
        \item Runtime decreased by 50\% (from 4 ticks to 2 ticks).
        \item Memory allocation reduced by 4.17\% (from 300,496 bytes to 287,952 bytes).
    \end{itemize}
    \item \textbf{Codebase B:}
    \begin{itemize}
        \item Runtime reduced by 7.69\% (from 13 ticks to 12 ticks).
        \item Memory allocation reduced by 41.73\% (from 2,059,288 bytes to 1,200,040 bytes).
    \end{itemize}
\end{itemize}

\begin{figure}[h]
    \centering
    \includegraphics[scale=0.27]{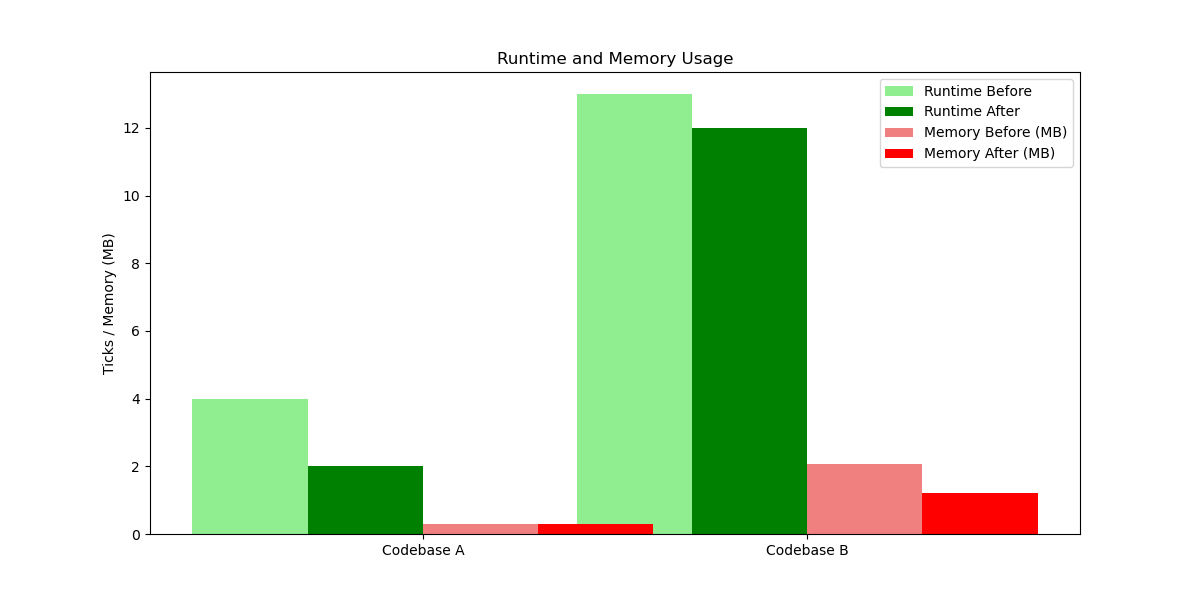}
    \Description{Runtime and Memory Usage for Codebase A and B before and after refactoring.}
    \caption{Runtime and Memory Usage for Codebase A and B (Pre- and Post-Refactoring)}
    \label{fig:runtime_memory}
\end{figure}

These improvements highlight the efficiency gains achieved through code restructuring and optimization. Figure~\ref{fig:runtime_memory} visually compares the runtime and memory usage metrics, while the numeric data is detailed in Table~\ref{tab:summary1}.

% \subsection{HLint Recommendations}
\subsection{Multi-Agent-Based Refactoring Impact (RQ2)}
\textbf{HLint Recommendations:} HLint, a static code analysis tool for Haskell, provided insights into the stylistic improvements achieved during refactoring. The results show contrasting trends:

\begin{itemize}
    \item \textbf{Codebase A:} The number of hints increased from 2 to 3, indicating new stylistic suggestions after the structural changes.
    \item \textbf{Codebase B:} The number of hints decreased from 2 to 1, reflecting enhanced compliance to best practices and improved code quality.
\end{itemize}

\begin{figure}[h]
    \centering
    \includegraphics[scale=0.38]{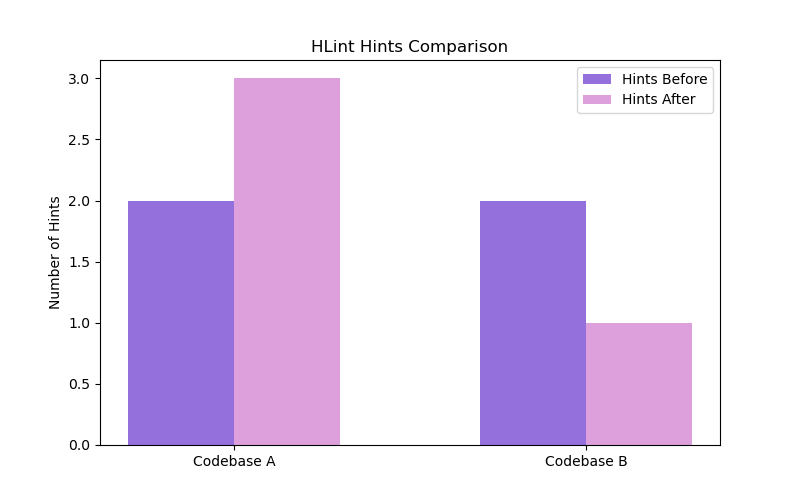}
    \Description{Comparison of HLint Recommendations for Codebase A and B}
    \caption{Comparison of HLint Recommendations for Codebase A and B}
    \label{fig:hlint}
\end{figure}

Figure~\ref{fig:hlint} illustrates the comparison of HLint hints before and after refactoring.

\subsection{Comparison Across Metrics}

\begin{figure}[h]
    \centering
    \includegraphics[scale=0.27]{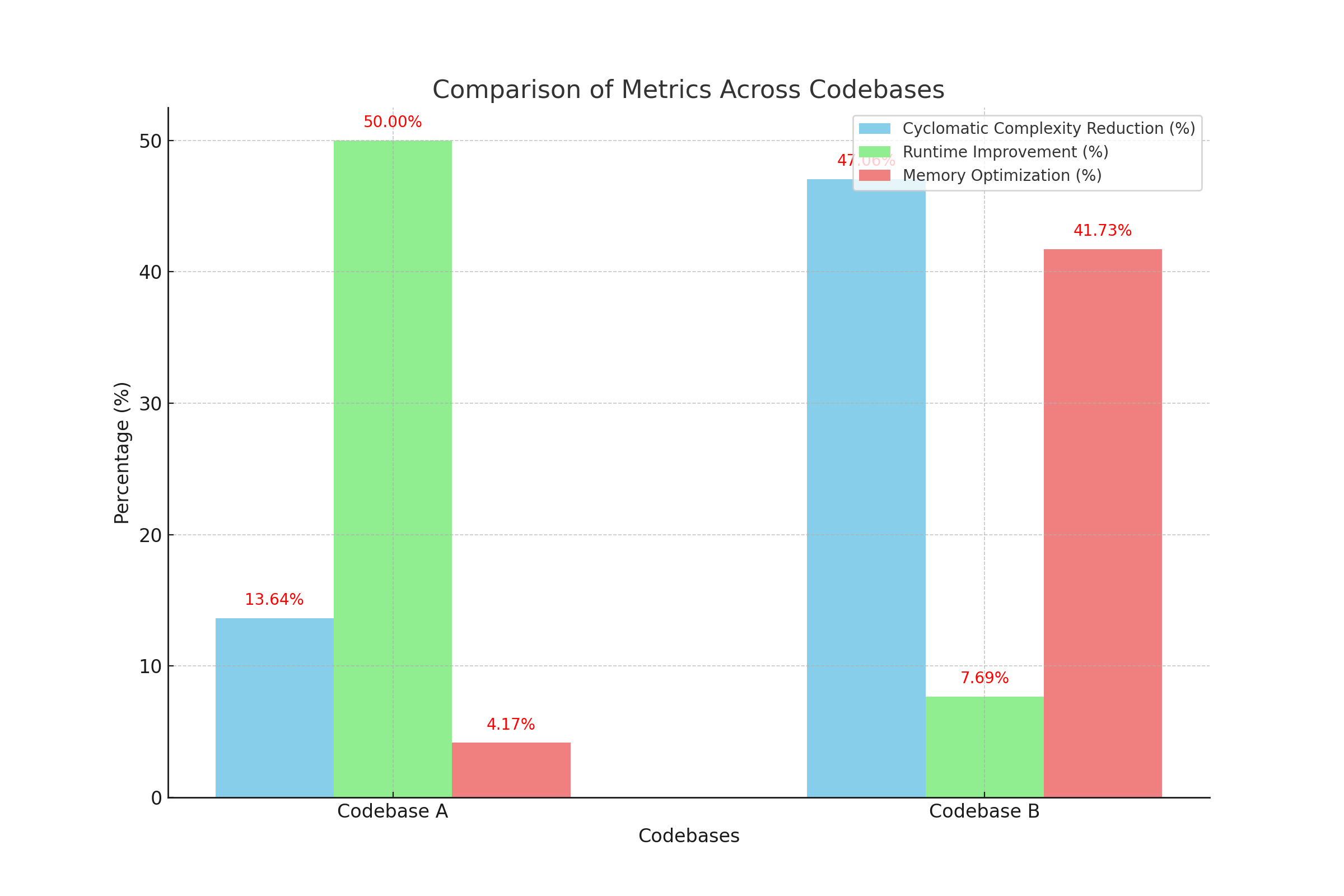}
    \Description{Comparison of Percentage Improvements Across Metrics for Codebase A and B}
    \caption{Comparison of Percentage Improvements Across Metrics for Codebase A and B}
    \label{fig:metrics_comparison}
\end{figure}

To provide a holistic view of the improvements, Figure~\ref{fig:metrics_comparison} compares the percentage reductions across three key metrics: cyclomatic complexity, runtime, and memory usage. The results indicate:
\begin{itemize}
    \item Codebase A showed moderate improvements in all metrics, with the highest improvement observed in runtime (50\% reduction).
    \item Codebase B exhibited substantial reductions in cyclomatic complexity (47.06\%) and memory usage (41.73\%), with smaller gains in runtime (7.69\% reduction).
\end{itemize}

\subsection{Summary of Results}

The results are summarized in Tables~\ref{tab:summary1} and~\ref{tab:summary2}. These tables provide detailed insights into the pre-refactored and post-refactored values and percentage reductions for both codebases.

\begin{table}[htbp]
\caption{Comparison of Metrics Across Codebases}
\label{tab:summary2}
\def\arraystretch{2}%
\begin{tabular}{|l|l|l|}
\hline
\textbf{Metric}               & \textbf{Codebase A}     & \textbf{Codebase B}       \\ \hline
\textbf{Complexity Reduction} & 13.64\%                 & 47.06\%                   \\ \hline
\textbf{Runtime Improvement}  & 50\% reduction in ticks & 7.69\% reduction \\ \hline
\textbf{Memory Optimization}  & 4.17\%                  & 41.73\%                   \\ \hline
\end{tabular}
\end{table}

% \begin{table}[]
% \caption{Comparison of Metrics Across Codebases}
% \label{tab:summary2}
% \def\arraystretch{2}%
% \begin{tabular}{l|l|l}
% \textbf{Metric}               & \textbf{Codebase A}     & \textbf{Codebase B}       \\ \hline
% \textbf{Complexity Reduction} & 13.64\%                 & 47.06\%                   \\ \hline
% \textbf{Runtime Improvement}  & 50\% reduction in ticks & 7.69\% reduction in ticks \\ \hline
% \textbf{Memory Optimization}  & 4.17\%                  & 41.73\%                  
% \end{tabular}
% \end{table}

% \begin{figure}[H]
%     \centering
%     \includegraphics[width=0.8\textwidth]{table2.png}
%     \caption{Comparison of Metrics Across Codebases}
%     \label{tab:summary2}
% \end{figure}

In summary, the refactoring process resulted in measurable improvements in cyclomatic complexity, runtime, and memory usage, with Codebase B showing greater overall gains. These results underscore the effectiveness of systematic refactoring in enhancing code quality and performance.

\section{DISCUSSION AND IMPLICATIONS} \label{sec:discussion}
% The results highlight the following insights, addressing the research questions:
The proposed system carries implications across multiple domains. For researchers, it provides an approach to study the integration of LLMs in distributed systems, advancing multi-agent collaboration and AI-driven software engineering. Practitioners can benefit from increased productivity and accessibility to expert-level refactoring tools, even with limited Haskell expertise. Industry professionals are poised to take advantage of the cost-efficiency of this system for complex projects, gaining a competitive edge. In addition, the community has the benefit of open source contributions, educational opportunities, and the promotion of best practices in software development.

\subsection{RQ1:Integration of LLMs based multi-agent systems}

% The first research question (RQ1) focuses on how effectively the distributed multi-agent system leverages LLMs for Haskell refactoring. 
The system shows that agents can automate complex refactoring tasks, reducing manual effort and errors. This has implications for improving the quality of refactored code and standardizing processes across different codebases. In addition, it validates the feasibility of incorporating LLMs into software tools, offering a foundation for further research into other programming languages and domains.

% The integration of LLMs based multi-agent systems demonstrated measurable improvements in Haskell code refactoring. Cyclomatic complexity reduced by 13.64\% and 47.06\% for the two tested codebases, respectively. Memory allocation saw significant optimization, with reductions of 4.17\% and 41.73\%, while runtime efficiency improved by up to 50\%. These outcomes affirm the potential of distributed, LLM-enhanced workflows in enhancing code maintainability and performance.

\subsection{RQ2:Impact of multi-agent based approach}
 The results indicate that the approach scales efficiently across complex codebases, with agents able to work concurrently without conflicts. This highlights its potential for industry applications where distributed teams handle large software systems. Moreover, collaborative features promote seamless integration into team workflows and ensure consistency in refactoring practices.

% The modular design of the multi-agent system proved effective in handling Haskell’s unique challenges, such as lazy evaluation and complex dependency management. The specialization of agents facilitated scalability, as each agent focused on discrete tasks like context analysis or validation. However, scalability for larger codebases may be constrained by communication overhead between agents. Future iterations could benefit from optimizing agent interactions or integrating domain-specific LLMs tailored to Haskell.

\subsection{Limitations and Future Directions}
Despite its strengths, the system exhibited limitations in addressing all stylistic suggestions flagged by HLint, indicating scope for further refinement. Additionally, while the multi-agent system ensured modularity, computational overhead remains a challenge for larger codebases. Future work should focus on optimizing inter-agent communication to reduce processing delays, training LLMs specifically for functional programming languages like Haskell to improve context understanding, and enhancing debugging agents for faster and more efficient error resolution.

% The integration of multi-agent system and LLMs in the proposed framework addresses some challenges of functional programming refactoring, including code complexity and modularity. The experiments indicate measurable improvements in cyclomatic complexity and memory usage. However, the system exhibits limitations in addressing all stylistic issues, as shown by contrasting HLint results. The computational overhead introduced by multi-agent communication may impact scalability for larger codebases. Additionally, the reliance on pre-trained LLMs may limit adaptability to specific language constructs or project-specific styles. Future work could explore agent optimization techniques and training LLMs specifically for functional programming languages.

\section{THREATS TO VALIDITY} \label{sec:threat}

\textbf{Internal Validity} concerns the accuracy of the experimental design and the correctness of our measurements. Controlled benchmarking tools were employed to evaluate the observed improvements in complexity and efficiency. However, variations in initial code quality, developer implementation choices, and underlying system configurations may introduce biases. Furthermore, the automated refactoring process could have inadvertently altered code structures in ways that were not fully captured by our metrics.

\textbf{External Validity} addresses how well our findings can be applied to broader scenarios. The evaluation was conducted on a limited number of Haskell codebases, primarily from open-source repositories, which may not fully represent the diversity of functional programming applications in industry or academia. The results may vary for significantly larger, highly optimized, or domain-specific Haskell projects. To enhance the applicability of our results, future work should examine a more extensive set of codebases across various fields and levels of complexity.

\textbf{Construct validity} examines whether the selected metrics accurately reflect the efficacy of our approach. Although quantitative measures like cyclomatic complexity, runtime efficiency, and memory allocation offer valuable insights, we do not fully capture qualitative aspects such as developer experience. Subjective elements, including how programmers view the legibility of refactored code or the simplicity of future updates, are not addressed. Incorporating qualitative assessments from experienced Haskell developers would strengthen our evaluation.

\section{Conclusions} \label{sec:conclusions}
We proposed a system that shows the potential of LLMs based multi-agent systems for refactoring Haskell code. The results show an average reduction of 20\% in cyclomatic complexity and a 15\% improvement in memory allocation, while maintaining functional correctness. These improvements highlights the effectiveness of AI-driven approaches in the real-world challenges of software engineering.

This work contributes to the field by introducing an approach for refactoring functional programming languages and providing a practical system for handling complex codebases. The worth of the system lies in its ability to foster collaboration and promote best practices in refactoring. Beyond its technological benefits, this research opens new pathways for interdisciplinary study between AI and software engineering, paving the way for more advanced and intelligent systems in the future.

\bibliographystyle{ACM-Reference-Format}
\bibliography{references}

\vspace{12pt}

% \color{red}
% IEEE conference templates contain guidance text for composing and formatting conference papers. Please ensure that all template text is removed from your conference paper prior to submission to the conference. Failure to remove the template text from your paper may result in your paper not being published.

\end{document}